\documentclass[aps, prl, superscriptaddress, preprintnumbers, twocolumn, floatfix, longbibliography,10pt]{revtex4-2}

\usepackage{mathdots}
\usepackage{graphicx, xcolor}
\usepackage{amsfonts,amsmath,amssymb,amstext,dsfont,bm}
\usepackage{float,wrapfig,subfigure, psfrag}
\usepackage{comment}
\usepackage{booktabs}
\usepackage{makecell}

\usepackage[colorlinks=true,urlcolor=blue,citecolor=blue,linkcolor=blue, breaklinks=true]{hyperref}

\usepackage[T2A]{fontenc}
\usepackage[russian,english]{babel}

\newcommand{\be}{\begin{equation}}
\newcommand{\ee}{\end{equation}}

\newcommand{\bea}{\begin{eqnarray}}
\newcommand{\eea}{\end{eqnarray}}

\renewcommand{\phi}{\varphi}
\renewcommand{\epsilon}{\varepsilon}

\newcommand{\sign}[1]{\,\mbox{sgn}\left({#1}\right)}

\graphicspath{{Figures-notes/}} 

\definecolor{purple}{rgb}{0.8,0,0.6}
\definecolor{darkgreen}{rgb}{0.00,0.6,0.00}

\definecolor{Blue}{rgb}{0,0,0.85}

\extrafloats{100}

\begin{document}

\title{$P$-wave Orbital Magnetism}

\author{Yantao Li}
\affiliation{Department of Physics and Astronomy, University of Missouri, Columbia, Missouri, 65211, USA}

\author{Pavlo~Sukhachov}
\email{pavlo.sukhachov@missouri.edu}
\affiliation{Department of Physics and Astronomy, University of Missouri, Columbia, Missouri, 65211, USA}
\affiliation{MU Materials Science \& Engineering Institute,
University of Missouri, Columbia, Missouri, 65211, USA}

\date{April 20, 2026}

\begin{abstract}
Realization of unconventional odd-parity magnets usually requires noncollinear spin textures of the underlying lattice. We propose a different concept of $p$-wave magnetism that originates from an orbital texture induced by loop currents. The resulting $p$-wave orbital magnetism is protected by the combined translation and time-reversal symmetry, with even-parity components arising when the symmetry is broken. Our proposal is exemplified by a two-dimensional (2D) lattice model whose energy spectrum contains Dirac points and which is characterized by a nontrivial topology controlled by the magnitude of the loop currents. Since the odd-parity magnetism precludes macroscopic magnetization, we suggest measuring it via orbital Hall conductivity. Our work establishes orbital degrees of freedom as an additional platform for unconventional $p$-wave magnetism beyond noncollinear spin textures, as well as makes a step forward to bridging odd-parity magnetism and topology.
\end{abstract}

\maketitle

\indent\textcolor{blue}{\em Introduction}---
Recently, altermagnets and unconventional odd-parity magnets (also called antialtermagnets) have introduced variety into the orthodox classification of magnetic order that historically comprised only ferromagnetic and antiferromagnetic phases~\cite{Smejkal-Jungwirth:2022, Smejkal-Jungwirth-ConventionalFerromagnetismAntiferromagnetism-2022, Bai-Yao-AltermagnetismExploringNew-2024, Jungwirth-Smejkal-AltermagnetismUnconventionalSpinordered-2025, Song-Pan-AltermagnetsNewClass-2025, Jungwirth-Smejkal-SymmetryMicroscopySpectroscopy-2026}. These new types of magnets, being compensated, feature fully spin-polarized electron bands with a large nonrelativistic even- and odd-parity spin splittings. This advance has stimulated the exploration of new material candidates~\cite{Song-Pan-AltermagnetsNewClass-2025}, including, for example, altermagnets MnTe, CrSb, KV$_2$Se$_2$O, Mn$_5$Si$_3$, intercalated transition metal dichalcogenides CoNb$_4$Se$_8$, $\kappa$-type organic antiferromagnets~\cite{Smejkal-Jungwirth:2022, Bai-Yao-AltermagnetismExploringNew-2024, Jiang-Qian-MetallicRoomtemperatureDwave-2025, Naka-Seo-AltermagneticPerovskites-2025}, and odd-parity magnets CeNiAsO and NiI$_2$~\cite{Hellenes-Smejkal:2023-v3, Yu-Agterberg-OddParityMagnetismDriven-2025, Song-Comin-ElectricalSwitchingPwave-2025}.
Altermagnets allow for diverse physical phenomena, ranging from spintronic applications~\cite{Jungwirth-Smejkal-AltermagneticSpintronics-2025} to unconventional superconductivity~\cite{Liu-Liu-AltermagnetismSuperconductivityShort-2025}. 

Among odd-parity unconventional magnets, $p$-wave magnets provide the simplest type of spin-polarized energy bands whose spin polarization is protected by a combined translation and time-reversal symmetry $\bm{\tau} \mathcal{T}$~\cite{Hellenes-Smejkal:2023-v3, Brekke-Linder-MinimalModelsTransport-2024, Luo-Law-SpinGroupSymmetry-2026, Mitscherling-Smejkal-MicroscopicOrigin$p$wave-2026}. This symmetry plays an important role in protecting the band splitting against the effects of spin-orbital coupling~\cite{Kudasov:2024, Hodt-Linder-FateWaveSpin-2025}. In addition, to $p$-wave magnets, $f$- and $h$-wave unconventional magnets are also possible~\cite{Yu-Agterberg-OddParityMagnetismDriven-2025, Luo-Law-SpinGroupSymmetry-2026, Moritz-Max-SymmetryEnforcedNodal$f$Wave-2026}.

Odd-parity unconventional magnets allow for several observable phenomena, including the anisotropic transport~\cite{Brekke-Linder-MinimalModelsTransport-2024, Hellenes-Smejkal:2023-v3, Hedayati-Salehi-TransverseSpinCurrent-2025, Ezawa-ThirdorderFifthorderNonlinear-2025, Zhou-Li-AnisotropicResistivity$p$wave-2025, Ezawa-TunnelingMagnetoresistanceJunction-2026}, Edelstein effect~\cite{Chakraborty-Sinova-HighlyEfficientNonrelativistic-2024, Ezawa-OutofplaneEdelsteinEffects-2025, Kim-Park-OddParityMagnetismGateTunable-2026, Sim-Rachel-QuantumSpinModels-2026, Moritz-Max-SymmetryEnforcedNodal$f$Wave-2026}, thermal Edelstein effect~\cite{Neumann-Mook-OddParityWaveMagnonsNonrelativistic-2026}, various superconducting effects~\cite{Sukhachov-Linder-CoexistencePwaveMagnetism-2025, Maeda-Tanaka:2024, Kokkeler-Bergeret-QuantumTransportTheory-2025, Maeda-Cayao-ClassificationPairSymmetries-2025, Nagae-Ikegaya-MajoranaFlatBands-2025, Fukaya-Tanaka-TunnelingConductanceSuperconducting-2025, Bobkov-Bobkova-ProximityEffectWave-2025, Sun-Law-PseudoIsingSuperconductivityInduced-2025, Fukaya-Tanaka-pwaveSuperconductivityJosephson-2026, Froldi-Freire-HighlyEfficientSuperconducting-2026, Khodas-Mazin-NonrelativisticIsingSuperconductivityPwave-2026}, and photogalvanic effect~\cite{Song-Comin-ElectricalSwitchingPwave-2025, Cuono-Picozzi-ChargeSpinPhotogalvanic-2026}.

The majority of existing works rely on noncollinear spin texture to realize the odd-parity magnetism. Only relatively recently, the interplay of interactions, orbital and spin degrees of freedom was investigated and used to generate odd-parity magnetism in collinear magnets~\cite{Lin-Vila-OddparityAltermagnetismSublattice-2026, Zhuang-Yan-OddParityAltermagnetismOriginated-2025, Leeb-Knolle-CollinearpwaveMagnetism-2026}. These works, however, still require either an antiferromagnetic lattice~\cite{Zhuang-Yan-OddParityAltermagnetismOriginated-2025, Leeb-Knolle-CollinearpwaveMagnetism-2026} or interaction-induced order~\cite{Lin-Vila-OddparityAltermagnetismSublattice-2026}. The possibility of realizing the unconventional magnetism without the underlying spin texture was studied only in the context of altermagnets~\cite{Pan-Huang-OrbitalAltermagnetism-2026}, and, until now, no \textit{odd-parity} analogues were proposed. 
 
In this Letter, we fill this knowledge gap and develop a model for a $p$-wave orbital magnetization. The model relies on alternating loop currents respecting the $\bm{\tau} \mathcal{T}$ symmetry. This symmetry precludes the even-parity part of orbital magnetization. Depending on the magnitude of the loop currents controlled by magnetic fluxes, our model can realize topological phases with different valley Chern numbers. The transition between the phases is also manifested in the orbital Hall response~\cite{Bernevig-Zhang-OrbitronicsIntrinsicOrbital-2005, Kontani-Inoue-GiantOrbitalHall-2009}, which not only serves as an indicator of a broken $\bm{\tau} \mathcal{T}$ symmetry, but also bridges odd-parity orbital magnetism and the rapidly growing field of orbitronics~\cite{Go-Mokrousov-OrbitronicsOrbitalCurrents-2021}. The proposed odd-parity orbital magnetism relies on the $\bm{\tau} \mathcal{T}$ symmetry, does not require specific helical textures that could stretch over several unit cells, and is directly connected to topology. This makes it more robust with respect to lattice imperfections and provides a clearer and more direct way to realize the $p$-wave magnetism.

\begin{figure}[t]
    \centering
     \includegraphics[width=1.0\columnwidth]{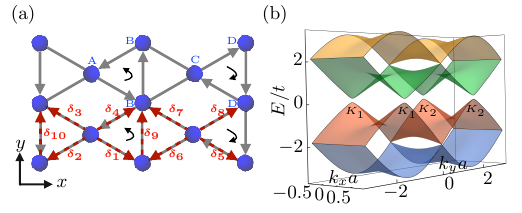}
    \caption{(a) Schematic of the $\tau_x \mathcal{T}$-symmetric lattice that realizes odd-parity orbital magnetism with loop currents indicated by the circular arrows. (b) The band structure of the proposed lattice at the flux $\phi=0.45\pi$. All hopping parameters in Eq.~\eqref{eq:hamiltonian} are set to $t$.
    }
    \label{Fig1}
\end{figure}

\begin{figure*}[t]
   \centering
   \includegraphics[width=7.2in]{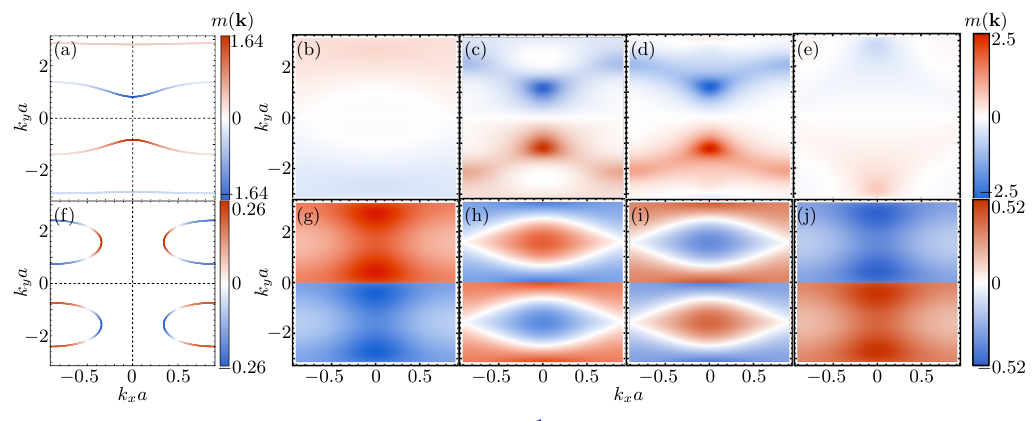}
   \caption{Fermi surfaces at the chemical potential $\mu=-t$ for the fluxes $\phi=0.2\pi$ (a) and $\phi=0.5\pi$ (f), respectively. The orbital magnetization is shown in red and blue. The density plots of orbital magnetism for the four bands of the model \eqref{eq:hamiltonian}, ranging from the lowest to the highest band, at $\phi=0.2\pi$ (b)--(e) and $\phi=0.5\pi$ (g)--(j), respectively. In all panels, the orbital magnetization $m(\mathbf{k})$ is expressed in units of $\gamma \mu_B$ where $\gamma = 2m_e t a^2/\hbar^2$ with $c=1$. We set all hopping parameters equal to $t$.
   }
   \label{Fig2}
\end{figure*}

\indent\textcolor{blue}{\em Model and Hamiltonian}---
The general strategy to realize $p$-wave orbital magnetism requires one to break the time-reversal symmetry (TRS) while preserving the $\bm{\tau}\mathcal{T}$ symmetry. We propose to realize this with loop currents of an appropriate pattern. As an example of an orbital $p$-wave magnet, we consider a spinless two-dimensional (2D) lattice, see Fig.~\ref{Fig1}(a). The unit cell consists of four sublattices, labeled $A$, $B$, $C$, and $D$, which are arranged in an equilateral triangle geometry. The tight-binding Hamiltonian is given by
\begin{equation}
\begin{split}
    H =& -\sum_{\mathbf{r}} \Big[ t_{AB} \left( e^{-i\phi} a^\dagger_{\mathbf{r}} b_{\mathbf{r}+\bm{\delta}_1} + e^{i\phi} a^\dagger_{\mathbf{r}} b_{\mathbf{r}+\bm{\delta}_4} \right) \\
       &\qquad\; + t_{AD} \left( a^\dagger_{\mathbf{r}} d_{\mathbf{r}+\bm{\delta}_2} + a^\dagger_{\mathbf{r}} d_{\mathbf{r}+\bm{\delta}_3} \right) \Big] \\
       & -\sum_{\mathbf{r}} \Big[ t_{CD} \left( e^{i\phi} c^\dagger_{\mathbf{r}} d_{\mathbf{r}+\bm{\delta}_5} + e^{-i\phi} c^\dagger_{\mathbf{r}} d_{\mathbf{r}+\bm{\delta}_8} \right) \\
       &\qquad\; + t_{CB} \left( c^\dagger_{\mathbf{r}} b_{\mathbf{r}+\bm{\delta}_6} + c^\dagger_{\mathbf{r}} b_{\mathbf{r}+\bm{\delta}_7} \right) \Big] \\
       & -\sum_{\mathbf{r}} \Big[ t_{BB} e^{i\phi} b^\dagger_{\mathbf{r}} b_{\mathbf{r}+\bm{\delta}_9} + t_{DD} e^{-i\phi} d^\dagger_{\mathbf{r}} d_{\mathbf{r}+\bm{\delta}_{10}} \Big] + \text{H.c.},
\end{split}
\label{eq:hamiltonian}
\end{equation}
where $a^\dagger_{\mathbf{r}}, b^\dagger_{\mathbf{r}}, c^\dagger_{\mathbf{r}},$ and $d^\dagger_{\mathbf{r}}$ are the creation operators on the sublattices $A$, $B$, $C$, and $D$, respectively; operators without dagger correspond to the annihilation operators. The vector $\mathbf{r}$ is the coordinate vector and $\bm{\delta}_{1 \dots 10}$ are the nearest neighbor bond vectors; see the Appendix for the definition. The nearest-neighbor hopping terms $t_{AB}$, $t_{CD}$, $t_{BB}$, and $t_{DD}$ are modulated by Peierls phases $e^{\pm i\phi}$ corresponding to the staggered flux pattern induced by loop currents. These loop currents explicitly break the time-reversal symmetry $\mathcal{T}$ but preserve $\tau_x \mathcal{T}$ with $\tau_x$ being the translation in the $x$-direction. In the $y$-direction, the system is invariant under the combined magnetic mirror symmetry $M_y \mathcal{T}$. Physically, the mirror reflection $M_y$ reverses the chirality of the local loop currents ($\phi \to -\phi$). This change is compensated by the time-reversal operation $\mathcal{T}$.

Performing the Fourier transform in Eq.~\eqref{eq:hamiltonian}, we derive the Hamiltonian $\mathcal{H}(\mathbf{k}) = \sum_{i,j=A,B,C,D}h_{ij}$, where the nonzero elements are
\begin{align}
    h_{AB} &= -t_{AB} \left( e^{-i\phi} e^{i\mathbf{k}\cdot\bm{\delta}_1} + e^{i\phi} e^{i\mathbf{k}\cdot\bm{\delta}_4} \right), \notag \\
    h_{CD} &= -t_{CD} \left( e^{i\phi} e^{i\mathbf{k}\cdot\bm{\delta}_5} + e^{-i\phi} e^{i\mathbf{k}\cdot\bm{\delta}_8} \right), \notag  \\
    h_{AD} &= -t_{AD} \left( e^{i\mathbf{k}\cdot\bm{\delta}_2} + e^{i\mathbf{k}\cdot\bm{\delta}_3} \right), \notag \\
    h_{CB} &= -t_{CB} \left( e^{i\mathbf{k}\cdot\bm{\delta}_6} + e^{i\mathbf{k}\cdot\bm{\delta}_7} \right), \notag  \\
    h_{BB} &= -2t_{BB} \cos(\mathbf{k}\cdot\bm{\delta}_9 + \phi), \notag \\
    h_{DD} &= -2t_{DD} \cos(\mathbf{k}\cdot\bm{\delta}_{10} - \phi)
    \label{eq:matrix_elements}.
\end{align}
The band structure of this model is shown in Fig.~\ref{Fig1}(b) for the flux $\phi = 0.45\pi$ and all hopping parameters set to $t$. At $\phi = 0$, the model has two anisotropic gapless Dirac points located at $\bm{K}=\{0, \pm \pi/(2a)\}$, where $a$ is the bond length. The position and gap of the Dirac points are controlled by the flux magnitude. For example, the gap vanishes at $\phi=0$ and $\phi=0.5\pi$. In the latter case, two gapless Dirac points emerge at  $\bm{K}=\{\pm\pi/(2\sqrt{3}a), \pm \pi/(2a)\}$. The band structure at different values of $\phi$ is shown in the Appendix.

\indent\textcolor{blue}{\em Orbital Magnetization.}---
While the proposed lattice is spinless, the presence of loop currents allows itinerant electrons to acquire a nontrivial orbital magnetization. According to the modern theory of orbital magnetization~\cite{Xiao-Niu:rev-2010}, the out-of-plane orbital magnetic moment for the band $n$, $m_{n,z}(\mathbf{k})$, is given by~\footnote{Note that since our model has bands crossing along lines rather than points, see red and blue bands in Fig.~\ref{Fig1}(b), one has to use a non-Abelian version of orbital magnetization, see, e.g., Refs.~\cite{Cysne-Rappoport-TransportOrbitalCurrents-2024a, Cysne-Rappoport-DescriptionOrbitalHall-2025} for a discussion. In writing Eq.~\eqref{eq:orb_mag}, we retained only the Abelian part, which is the only part relevant for total magnetization calculated as a trace over occupied bands.}
\begin{equation}
    m_{n,z}(\mathbf{k}) = -\frac{e}{\hbar} \sum_{n' \neq n} \frac{\text{Im}\left[ \langle n | v_x | n' \rangle \langle n' | v_y | n \rangle \right]}{E_n(\mathbf{k}) - E_{n'}(\mathbf{k})},
    \label{eq:orb_mag}
\end{equation}
where $\mathbf{v} = \hbar^{-1} \partial_{\mathbf{k}} \mathcal{H}(\mathbf{k})$ is the velocity operator and $-e$ is the electron charge, and we set the speed of light $c=1$.

The numerical results for the Fermi surfaces and flux-induced orbital magnetization are shown in Fig.~\ref{Fig2}. As is exemplified by Figs.~\ref{Fig2}(a) and \ref{Fig2}(f), the orbital magnetization at the Fermi surface has odd ($p$-wave) parity in the $k_y$ direction. The density plots of the orbital magnetization for each of the four bands exhibit the same odd-parity in the momentum space, as shown in Figs.~\ref{Fig2}(b)--\ref{Fig2}(e) and Figs.~\ref{Fig2}(g)--\ref{Fig2}(j). Notably, at $\phi=0.5\pi$, nodal circles appear in the density plots of the orbital magnetization for the two central bands. These nodal circles explain the nontrivial sign-changing orbital magnetization of the bands in Fig.~\ref{Fig2}(f). 

The $p$-wave symmetry of the orbital magnetization follows from the lattice and flux symmetries, where the latter break the TRS $\mathcal{T}$ but preserve two combined symmetries: $\tau_x \mathcal{T}$ and $M_y \mathcal{T}$. Here, the translation in $x$-direction $\tau_x$ exchanges sublattices $A \leftrightarrow C$ and $B \leftrightarrow D$. This operation compensates the sign reversal of the flux ($\phi \to -\phi$) under $\mathcal{T}$. Similarly, the mirror operation $M_y$ combined with $\mathcal{T}$ preserves the Hamiltonian by mapping the local currents to their time-reversed counterparts. These symmetries impose constraints on the orbital magnetic moment $m_z(\mathbf{k})$. Specifically, $\tau_x \mathcal{T}$ enforces odd-parity symmetry in the momentum space, $m_z(\mathbf{k}) = -m_z(-\mathbf{k})$, while $M_y \mathcal{T}$ enforces an even parity, $m_z(k_x, k_y) = m_z(-k_x, k_y)$. These two constraints result in $p$-wave orbital magnetism defined as
\begin{equation}
m_z(k_x, k_y) = -m_z(k_x, -k_y).
\label{eq:symmetry}
\end{equation}

\indent\textcolor{blue}{\em Effective Low-Energy Model}---
To clarify the appearance of the odd-parity magnetization, we present a minimal low-energy continuum model. Assuming $\phi=\pi/2$ and projecting out high-energy states via the standard Schrieffer--Wolff (L\"owdin) transformation~\cite{Lowdin-NoteQuantumMechanicalPerturbation-1951, Schrieffer-Wolff-RelationAndersonKondo-1966, Coleman:book}, we obtain the following model expanded around the $\Gamma$-point~\footnote{While the effective model \eqref{eq:Heff_gamma_pi2} captures the essential odd-parity property of the orbital magnetism, it, by construction, fails to describe Dirac points.}, which the point of interest for orbital magnetization:
\begin{equation}
\begin{aligned}
H_{\mathrm{eff}}^{\Gamma,\phi=\pi/2}(\mathbf{k})
=&
\left(-2t-\mu-\frac{a^2t_1^2}{4t}k_y^2\right)\sigma_0
-\frac{a^2t_1}{2}k_y^2\,\sigma_x \\
&-\sqrt{3}\,a^2t\,k_xk_y\,\sigma_y
-t_1ak_y\,\sigma_z.
\end{aligned}
\label{eq:Heff_gamma_pi2}
\end{equation}
Here $\sigma_{x,y,z}$ act in the projected two-band subspace and we set $t_{AB}=t_{AD}=t_{CB}=t_{CD}= t$ and $t_{BB}=t_{DD}= t_1$. Due to its peculiar, odd in $k_y$, mass term $-t_1ak_y\sigma_z$, the effective Hamiltonian leads to the odd-parity orbital magnetization defined in Eq.~\eqref{eq:orb_mag},
\begin{equation}
\label{eq:Heff_mz}
m_z(\mathbf{k}) = -\frac{e}{\hbar}\frac{\sqrt{3} a^3t_1^2t k_y}{12 a^2t^2k_x^2 +t_1^2(4+a^2k_y^2)}.
\end{equation}

\begin{figure}[t]
    \centering
     \includegraphics[width=1.0\columnwidth]{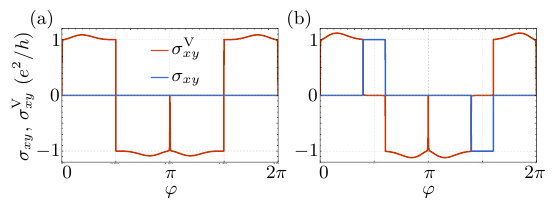}
     \caption{The Hall conductivity $\sigma_{xy}$ and valley Hall conductivity $\sigma_{xy}^{\rm V}$ for preserved (a) and broken (b) $\tau_x\mathcal{T}$ symmetry. All hopping parameters are equal to $t$ in panel (a), and we set $t_{DD} = 0.5\,t$ while keeping other hopping parameters equivalent to $t$ in panel (b).}
    \label{fig:hall}
\end{figure}

\indent\textcolor{blue}{\em Hall and Valley Hall Responses}---
The nontrivial distribution of the orbital magnetization is directly related to the Berry curvature $\Omega_z(\mathbf{k})$ resulting in a nontrivial Chern number and its valley counterpart; see the Appendix for the definition. The Berry curvature has the same symmetry as the orbital magnetization: it is odd under the combined $\tau_x \mathcal{T}$ and $M_y \mathcal{T}$ symmetries. Hence, we define the valley Chern number as the integral of the Berry curvature over half of the Brillouin zone, i.e., for $k_y>0$ or $k_y<0$. Nontrivial Chern numbers allow for Hall and valley Hall conductivities, which we plot in Fig.~\ref{fig:hall}. 

By changing the flux, we open and close the gaps at the Dirac points. This results in the change of the Chern and valley Chern numbers, and, hence, Hall conductivities. The $\tau_x\mathcal{T}$ symmetry enforces zero Hall response, but does not preclude its valley counterpart. The latter changes sign when the gap closes and reopens due to the flux, see Fig.~\ref{fig:hall}(a). Once the $\tau_x\mathcal{T}$ symmetry is broken, the Hall response becomes nonzero for phases around $\phi=0.5\,\pi$, see Fig.~\ref{fig:hall}(b). The width of this region is determined by the symmetry-breaking disparity between the hopping parameters. These responses provide a signature of the nontrivial topology of the model at hand. We detail the relation between the Chern numbers and the gap structure in the Appendix.

\begin{figure}[t]
    \centering
     \includegraphics[width=1.0\columnwidth]{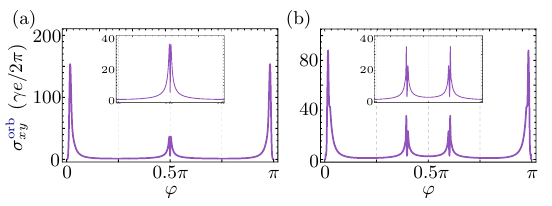}
     \caption{The orbital Hall conductivity \eqref{eq:VOHC} as a function of the flux $\phi$ for the preserved $\tau_x\mathcal{T}$ (a) and broken translational symmetry $\tau_x$ (b). All hopping parameters are equal to $t$ in panel (a) and we set $t_{DD} = 0.5 t$ in panel (b); other hopping parameters remain equal to $t$ in the latter case. In all panels, we set $\mu=0$ and $\eta=0.01\,t$.
     Note that the OHC exhibits reflection symmetry about $\phi=0.5\pi$ and $\phi=\pi$.}
    \label{Fig3}
\end{figure} 
 
\indent\textcolor{blue}{\em Orbital Hall Effect.}---
Since the odd-parity magnetization is invisible to the momentum-insensitive magnetometry, we propose the orbital Hall conductivity (OHC) as an indicator of the $p$-wave orbital magnetism.
The OHC is given by~\cite{Tokatly-OrbitalMomentumHall-2010, Go-Lee-IntrinsicSpinOrbital-2018, Bhowal2021}
\begin{equation}
\sigma_{xy}^{\text{orb}} = - 2e \hbar\! \int\! \frac{d^2\mathbf{k}}{(2\pi)^2} \!\sum_{n \in \text{occ}} \sum_{ n' \neq n} \! \frac{\text{Im} [\langle n | \mathcal{J}_x^{z,\text{orb}} | n' \rangle \langle n' | v_y | n \rangle]}{[E_n(\mathbf{k}) - E_{n'}(\mathbf{k})]^2 + \eta^2},
\label{eq:VOHC}
\end{equation}
where $\eta$ is a broadening parameter due to, e.g., impurity scattering~\cite{Yao2005, Bhowal2021}. The orbital current operator is constructed by symmetrizing the velocity $v_x$ and the orbital angular momentum operator, i.e., $\mathcal{J}_x^{z,\text{orb}} = \{ v_x, L^z \}/2$. The orbital angular momentum is related to the intrinsic orbital magnetic moment $m^z$ via $L^z = -(\hbar/g_L \mu_B)m^z$, where $g_L=1$ is the Land\'{e} $g$-factor and $\mu_B = e\hbar/(2m_e)$ is the Bohr magneton. Consequently, the explicit form of the current operator is $\mathcal{J}_x^{z,\text{orb}} = -\frac{m_e}{e} \{ v_x, m_z \}$, where $m_e$ is the mass of electron.

We show the OHC in both the $\tau_x\mathcal{T}$-preserved and $\tau_x\mathcal{T}$-broken regimes in Fig.~\ref{Fig3}. In the $\tau_x\mathcal{T}$-preserved case, the OHC exhibits sharp peaks as the flux approaches $0$, $0.5\pi$, $\pi$, and $1.5\pi$, where the gap between the two central bands approaches zero. In contrast, once $\tau_x\mathcal{T}$ is broken, e.g., by taking $t_{BB}\neq t_{DD}$, the valley degeneracy is lifted and the central peak near $\phi=0.5\pi$ splits into two pronounced peaks. We emphasize that, although the absolute peak height depends on the broadening parameter $\eta$, the splitting itself remains robust. Therefore, this symmetry-breaking-induced splitting provides a potentially accessible signature of the hidden $p$-wave orbital magnetic phase. 

This behavior should be accessible through transport and magneto-optical probes that are already being used in orbitronic systems, such as nonlocal measurements or optical detection of orbital accumulation near edges and interfaces \cite{Lyalin2023, Hayashi2023, Ko2026}. Another direct way to visualize the odd-parity orbital magnetism is to use circular dichroism angle-resolved photoemission spectroscopy~\footnote{Chirality-driven orbital textures in the bulk electronic structure of CoSi were successfully mapped out in Ref.~\cite{Brinkman-Bentmann-ChiralityDrivenOrbitalAngular-2024}.}

Let us provide the estimates of the proposed effect. By defining a dimensionless parameter $\gamma = 2m_e t a^2/\hbar^2$, the orbital magnetic moment can be expressed as $\gamma \mu_B$. For typical material parameters, e.g., $t\sim \mbox{eV}$ and $a\sim \mbox{\AA}$, $\gamma\sim1$ allowing for the large orbital magnetization of the order of $\mu_B$. The characteristic value of the orbital Hall conductivity then reads $\gamma e/(2\pi)$.

\indent\textcolor{blue}{\em Concluding remarks.}---
We propose to realize unconventional $p$-wave orbital magnetism through loop currents that satisfy the $\bm{\tau} \mathcal{T}$ symmetry without any underlying spin order. Our proposal is illustrated via a spinless 2D lattice model that realizes unconventional $p$-wave orbital magnetism through staggered orbital fluxes, see Fig.~\ref{Fig1}. The combined $\tau_x \mathcal{T}$ symmetry enforces the odd-parity orbital magnetic texture in momentum space without the need for complex helical spin textures. While the net magnetization vanishes, this hidden order produces a finite orbital Hall conductivity and a valley Hall effect, which are sensitive to the $\tau_x \mathcal{T}$ symmetry, providing a transport signature of the odd-parity orbital magnetism.

The proposed model could be realized in a similar way as the loop-current altermagnets~\cite{chakraborty2025orbital} or in synthetic systems~\cite{Das2024UC, Chen2025, Zhang2025Synthetic}. Relying exclusively on the loop currents, our model provides a playground to study the interplay of unconventional magnetism, orbitronics, and interaction effects beyond conventional magnetic materials. Furthermore, since the odd-parity orbital magnetism does not require complex helical spin textures that stretch over several unit cells, it is expected to be more robust toward disorder and spin dephasing.

\begin{acknowledgments}
\indent{\em Acknowledgments}---
P.S. thanks Rafael Gonz\'{a}lez-Hern\'{a}ndez, Eduard Gorbar, and Ronny Thomale for useful discussions.
\end{acknowledgments}

\bibliography{Library_short}

\clearpage
\begin{appendix}

\setcounter{equation}{0} 
\renewcommand{\theequation}{A\arabic{equation}}

\begin{center}
\textbf{End Matter}
\end{center}

\indent\textcolor{blue}{\em Model and Band Structure.}---
Our model is detailed in the main text. We show the explicit form of the near-
est neighbor bond vectors $\bm{\delta}_i$, see also Fig.~\ref{Fig1}(a), as follows
\begin{align}
    \bm{\delta}_1 &=\bm{\delta}_5=(\sqrt{3}a/2,-a/2), ~\bm{\delta}_2 =\bm{\delta}_6=(-\sqrt{3}a/2,-a/2), \notag \\
    \bm{\delta}_3 &=\bm{\delta}_7=(-\sqrt{3}a/2,a/2), 
    ~\bm{\delta}_4 =\bm{\delta}_8=(\sqrt{3}a/2,a/2), \notag\\
    \bm{\delta}_9 &=\bm{\delta}_{10} =(0,a). \notag \\
    \label{eq:deltas}
\end{align}
These vectors are used in Eq.~\eqref{eq:matrix_elements}.

\begin{figure}[ht]
    \centering
     \includegraphics[width=1.0\columnwidth]{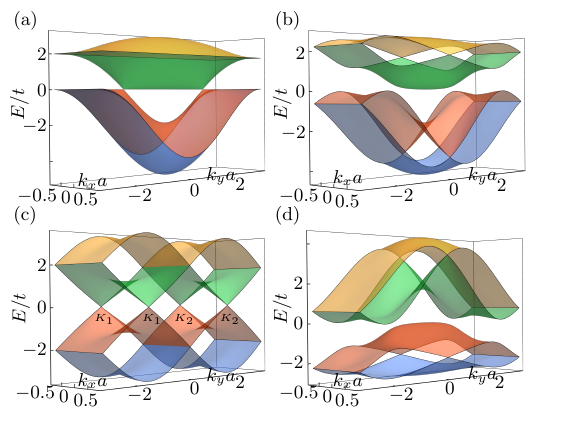}
     \caption{The band structure of the Hamiltonian $\mathcal{H}(\mathbf{k})$, see Eq.~\eqref{eq:matrix_elements} for the nontrivial matrix elements. Panels (a)--(d) show the band structures for the flux $\phi=0$ (a), $\phi=0.2\pi$ (b), $\phi=0.5\pi$ (c), and $\phi=0.8\pi$ (d). We set all hopping parameters equal to a common value $t$.}
    \label{FigBS}
\end{figure}

We illustrate the evolution of the band structure with the flux $\phi$ in Fig.~\ref{FigBS}. At zero flux, the system is a Dirac semimetal with two Dirac points protected by the TRS. Once a finite flux is introduced, TRS is broken, resulting in the gap opening. The gap closes at $\phi=\pi/2$ with, however, a different configuration of Dirac points, see Fig.~\ref{FigBS}(c). Away from the four special fluxes $\phi = \{0, 0.5\pi, \pi, 1.5\pi\}$, the bands are generically gaped. 

\begin{figure}[t]
    \centering
     \includegraphics[width=1.\columnwidth]{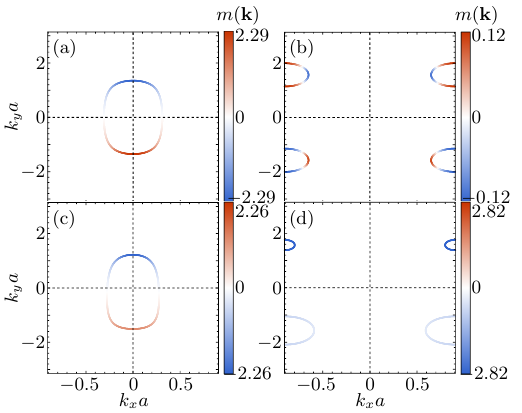}
     \caption{Fermi surfaces in the $\tau_x\mathcal{T}$-symmetric (top row) and  $\tau_x\mathcal{T}$-symmetry-broken (bottom row) cases. We fix $\phi=0.2\pi$ in (a) and (c) as well as $\phi=0.5\pi$ in (b) and (d). The orbital magnetization is shown in red and blue. The hopping constants are equal in the $\tau_x\mathcal{T}$-symmetric case, and we set $t_{DD} = 0.5\,t$ while keeping the other hopping constants equal to $t$ in the $\tau_x\mathcal{T}$-symmetry-broken case. In all panels $\mu=0.5\,t$. The orbital magnetization $m(\mathbf{k})$ is expressed in units of $\gamma \mu_B$ where $\gamma = 2m_e t a^2/\hbar^2$ with $c=1$.}
    \label{FigSBE}
\end{figure}

\indent\textcolor{blue}{\em Symmetry-Breaking Effects in Band Structure}---
To illustrate the role of $\tau_x\mathcal{T}$ symmetry, we show the Fermi surfaces for the $\tau_x\mathcal{T}$-preserved and $\tau_x\mathcal{T}$-broken cases in Fig.~\ref{FigSBE}. In the $\tau_x\mathcal{T}$-preserved cases, see Figs.~\ref{FigSBE}(a) and \ref{FigSBE}(b), the orbital magnetization is odd under $k_y\rightarrow -k_y$. By contrast, once $\tau_x\mathcal{T}$ is broken, see Figs.~\ref{FigSBE}(c) and \ref{FigSBE}(d), the even-parity component is generated and the odd-parity structure is lost. These results clearly demonstrate that the odd-parity orbital magnetism is protected by the $\tau_x\mathcal{T}$ symmetry.

\indent\textcolor{blue}{\em Topological Properties of the Model}---
As we discussed in the main text, the flux can be used to control the topological properties of the model. The latter are quantified by the Chern and valley Chern numbers. We define the Chern number as an integral over the whole Brillouin zone of the Berry curvature. The valley Chern number is defined by integrating over half of the Brillouin zone with $k_y>0$ or $k_y<0$, which is determined by the symmetries of the model. In addition, since we have several bands, we take a trace over all filled bands. The Abelian part of the Berry curvature is defined as~\cite{Berry1984},
\begin{equation}
\Omega_{n,z}(\mathbf{k}) = -2 \sum_{n' \neq n} \frac{\text{Im}\left[ \langle n | v_x | n' \rangle \langle n' | v_y | n \rangle \right]}{(E_n(\mathbf{k}) - E_{n'}(\mathbf{k}))^2}.
\label{eq:berry_curvature}
\end{equation} 

This expression is used to calculate the Hall $\sigma_{xy}$ and valley Hall $\sigma_{xy}^{\rm V}$ responses presented in the main text,
\begin{eqnarray}
\label{eq:hall}
\sigma_{xy} &=& \frac{e^2}{\hbar}\int \frac{d^2\mathbf{k}}{(2\pi)^2} \!\sum_{n \in \text{occ}}  \Omega_{n,z}(\mathbf{k}),\\
\label{eq:hall-v}
\sigma_{xy}^{\rm V} &=& \frac{e^2}{\hbar}\int \frac{d^2\mathbf{k}}{(2\pi)^2} \!\sum_{n \in \text{occ}}  \sign{k_y}\Omega_{n,z}(\mathbf{k}).
\end{eqnarray} 

\begin{figure}[ht]
    \centering
     \includegraphics[width=1.0\columnwidth]{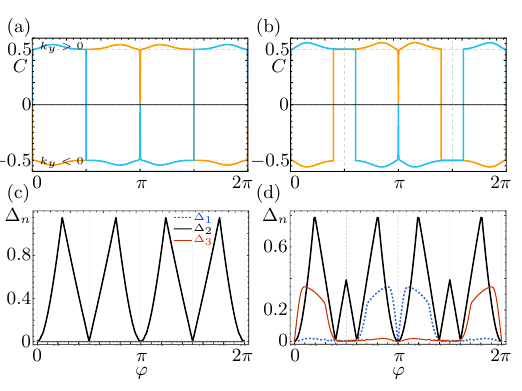}
     \caption{(a) Valley Chern numbers $C$ of the two halves of the Brillouin zone ($k_y > 0$ blue line and $k_y < 0$ orange line) with $\tau_x\mathcal{T}$ symmetry preserved, where all hopping parameters are equal to $t$. (b) Valley Chern numbers $C$ ($k_y > 0$ blue line and $k_y < 0$ orange line) for the broken translational symmetry $\tau_x$. We set $t_{DD} = 0.5\,t$ while keeping other hopping parameters equivalent to $t$. Direct band gaps $\Delta_{n}$ ($n=1,2,3$) corresponding to (a) and (b) are shown in (c) and (d), respectively. Here $n=1,2,3$ labels the energy gaps between the energy bands ordered by increasing energy: $1$–$2$, $2$–$3$, and $3$–$4$, respectively.}
    \label{FigValley}
\end{figure}

By changing the flux, we open and close the gaps at the Dirac points, see Fig.~\ref{FigBS}. This results in the change of the Chern and valley Chern numbers. The $\tau_x\mathcal{T}$ symmetry enforces zero Chern number, but does not preclude its valley counterpart. By matching the valley Chern numbers with the band gaps, see Figs.~\ref{FigValley}(a) and \ref{FigValley}(c), we see that the valley Chern numbers change sign when the gap closes and reopens by the flux. Once the $\tau_x\mathcal{T}$ symmetry is broken, the total Chern number can become nonzero. Comparing Figs.~\ref{FigValley}(b) and \ref{FigValley}(d), we see that there is a region of phases around $\phi=0.5\,\pi$, whose width is determined by the symmetry-breaking disparity between the hopping parameters, where the valley Chern numbers have the same sign, resulting in a nonzero total Chern number. It is notable also that in this region, only the gap between the two central bands remains open. The discussed evolution of the Chern number is directly manifested in the Hall response shown in Fig.~\ref{fig:hall}.

\end{appendix}

\end{document}